\documentclass[intlimits,twoside,a4paper]{article}

\usepackage{amsmath,amssymb}
\usepackage{graphicx}
\usepackage{wrapfig}

\usepackage[T2A]{fontenc}
\usepackage[cp1251]{inputenc}

\usepackage{epsf,multicol,ifthen}
\usepackage[eqsecnum]{cmpj3}
\usepackage{epstopdf}

\issue{2021}{24}{1}{13401}
\doinumber{10.5488/CMP.24.13401}

\title[The influence of adsorbed atoms concentration on the temperature coefficient of resonant frequency]%
{The influence of adsorbed atoms concentration on the temperature coefficient of resonant frequency of the quasi-Rayleigh wave}
\author[M.Ya.~Seneta \textsl{et al.}]{M.Ya.~Seneta\refaddr{1,2}\footnote{e-mail: marsen18@i.ua},
        R.M.~Peleshchak\refaddr{1,2}, A.I.~Nesterivskyi\refaddr{1}, N.I.~Lazurchak\refaddr{1}, S.K.~Guba\refaddr{2}}
\addresses{
\addr{1} Drohobych Ivan Franko State Pedagogical University, 24 I.Franko Str., 82100 Drohobych, Ukraine \\ 
\addr{2} Lviv Polytechnic National University, 12 Bandera street, 79013 Lviv, Ukraine
}

\date{Received March 3, 2020, in final form July 30, 2020}

\begin{document}

\maketitle

\begin{abstract}
Within the model of self-consistent connection of quasi-Rayleigh wave with adsorbed atoms, a method of constructing a new class of radiometric sensors of the temperature and concentration of adsorbed atoms  on the surface acoustic waves is proposed. Based on the developed theory of the dispersion and acoustic mode width of a quasi-Rayleigh wave on the adsorbed surface of monocrystals with a Zinc blende structure, the temperature coefficient of the resonant frequency of the surface acoustic wave is calculated depending on the temperature and on the  concentration of adsorbed atoms.

\end{abstract}

\section{Introduction}
Currently, the surface acoustic wave (SAW) devices are widely used in acoustoelectronics. In order to create a new class of microelectronic and nanoelectronic SAW devices with controlled parameters, such as sensors of temperature, sensors of the concentration of adsorbed atoms,  pressure sensors, gas sensors, biosensors, filters, signal processing units, automated actuators, etc., and to  develop radio measuring remote systems for control and measurement of physical values, for control of technological processes and the parameters of the operation of various mechanisms for identification of objects, it is necessary to investigate the regularities of the changes of the dispersion of surface acoustic waves, the width of the surface elastic acoustic mode on the surface of semiconductors, the mechanisms of the electron states excitation on the dynamically deformed adsorbed surface and on the border of semiconductors, depending on the concentration and on the type of adsorbed atoms. These studies will be used in SAW technologies for the production of wireless, passive, and hybrid multifunctional SAW sensors.

The use of surface acoustic waves in electronic devices was presented in many modern works~\cite{Liu16,Bor13,Zho16,Dev17,Mul16}. In the work~\cite{Liu16}, the operating principles of devices on the surface acoustic waves were described. In particular, the correspondence between the sound speeds and the temperature coefficients of frequency for traditionally used materials in SAW-devices was considered. In the work~\cite{Bor13}, the design and fabrication procedure was presented for SAW-based temperature, pressure and impedance sensors. The authors of~\cite{Bor13}  obtained a linear relationship between the resonance frequency shifts and the applied pressure and temperature as well as between the reflection coefficient and the changes of the impedance of the reflectors. The authors of work~\cite{Zho16}  made corresponding transformations of the dispersion equation of the film SAW-waveguide, using the experimental dependence of the phase SAW velocity on the basis of the ZnO film and its approximation. Based on the obtained analytical ratios and numerical results, the design parameters of the waveguide structures for providing a single-mode regime of SAW propagation were determined in~\cite{Zho16}. The review of the main studies of Rayleigh waves sensors while sounding the chemicals in the gaseous phase was conducted in the work~\cite{Dev17}. The authors emphasize that the use of new piezoelectric substrates, taking into account the acoustoelectronic effect, is a potential area of chemical SAW-sensors research.

However, the above theories describe the processes on the surface of a solid without taking into account the interaction of adsorbed atoms with SAW. These theories do not consider the regularities of the influence of the concentration of atoms adsorbed on the surface of the substrate on the surface characteristics.

Adsorption on the surface of the material was considered in the works~\cite{Evy16,Liu17,Cui15,Abb15,Don16,Li 15, May15}. In particular, in the work~\cite{Liu17} there were performed  calculations for  the study of the behavior of adsorbed environmental gas molecules on single-layer blue phosphorus, such as O$_{2}$, NO, SO$_{2}$, NH$_{3}$, H$_{2}$O, NO$_{2}$, CO$_{2}$, H$_{2}$S, CO and N$_{2}$, and for the study of their effect on electronic properties of the material of gas sensors. According to the theoretical estimates, chemical sensors based on multilayer black phosphorus were developed in the experiments and show good sensitivity not only to NO$_{2}$, NH$_{3}$ and H$_{2}$ gases~\cite{Cui15,Abb15,Don16}, but also to metal ions and humidity~\cite{Li 15, May15}.

Sensors based on typical semiconductors (including binary type A$_{3}$B$_{5}$, A$_{2}$B$_{6}$) are capable of detecting a large number of different substances at low temperatures, up to room temperature, which makes them quite promising. The operation of semiconductor sensors is based mainly on the relationship of the electrical conductivity of the near-surface layer of the spatial charge of the semiconductor with the charging of its surface due to the adsorption of foreign atoms~\cite{Kir20}.

However, in these works the influence of the mechanism of interaction of adatoms with a surface acoustic wave on the regularities of the changes of characteristics of SAW depending on the  concentration of adatoms is not investigated.

The purpose of this work is to investigate the influence of the concentration of adsorbed atoms on the temperature coefficient of resonant frequency of a surface acoustic wave within a self-consistent interconnection of a quasi-Rayleigh wave with adsorbed atoms on the surface of semiconductors having a Zinc blende structure.

\section{Formulation of the problem}
\begin{figure} [!b]
   \begin{center}
   \includegraphics[width=100mm]{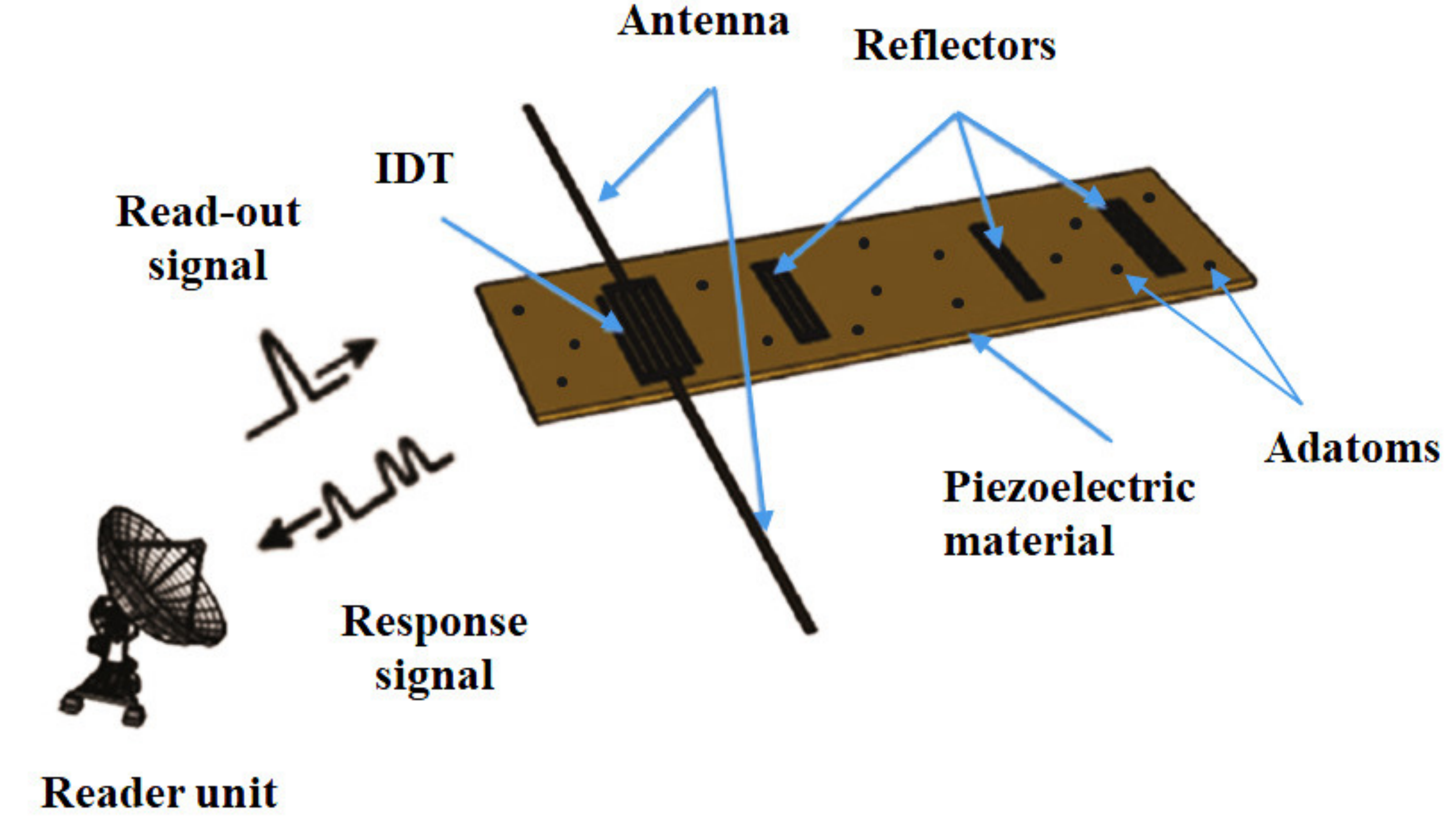}
   \caption{(Colour online) Schematic of the operating principle of the radiometric sensor of temperature and concentration of adsorbed atoms based on the SAW.}
   \label{fig1}
   \end{center}
\end{figure}
A fundamental factor that determines the advantages of acousto-electronic devices based on the SAW, including sensors, and in particular the advantages over the analogs built on other physical principles,  is the possibility of influence (action) on the SAW on the way of their propagation on the surface of a solid. The other important factor is that the velocity of propagation of acoustic waves is $10^5$ times lower than the velocity of propagation of electromagnetic waves.

Currently, based on surface acoustic waves, there are sensors of temperature, pressure, humidity,  as well as sensors that are capable of detecting  any changes in the electric field on the surface~\cite{Mul16}.

In this work, the construction of a new class of SAW-based radiometric sensors of temperature and concentration of adsorbed atoms (figure~\ref{fig1}) is based on the self-consistent connection between the quasi-Rayleigh wave and adsorbed atoms. Within this connection,   the regularity of the change of dispersion of quasi-Rayleigh wave and the width of the surface elastic acoustic wave depending on the concentration of the adsorbed atoms is investigated~\cite{Pel18}.

\begin{figure} [!t]
   \begin{center}
   \includegraphics[width=140mm]{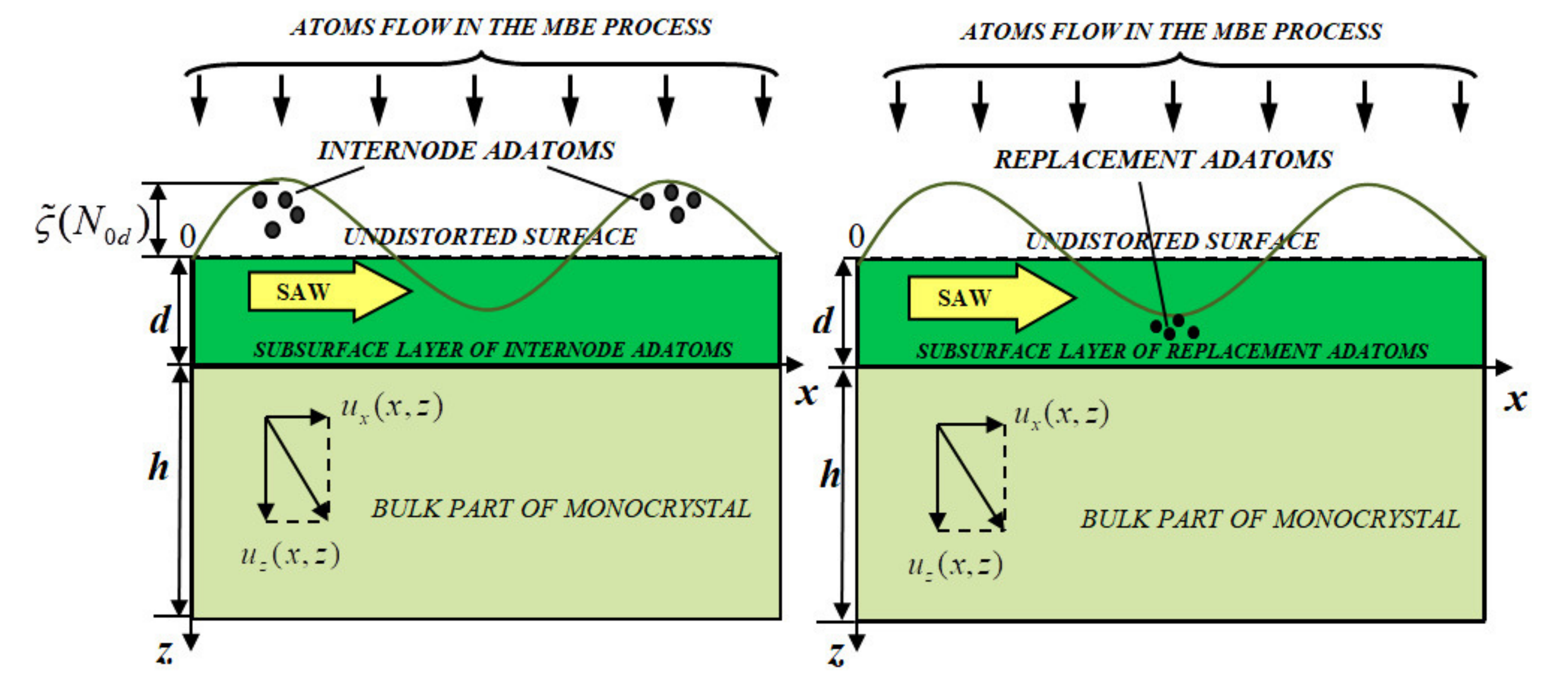}
   \caption{(Colour online) Model of the rough surface deformed by a surface acoustic wave and by the adsorbed atoms.
Figure~2a corresponds to the location of the adsorbed atoms in the internodes of the crystal lattice (adatoms are gathered in the regions of maximum tension deformation).
Figure~2b corresponds to the location of the adsorbed atoms at the nodes of the surface crystal lattice in which the covalent-ion radius is smaller than the covalent-ion radius of the surface crystalline lattice atom (adatoms are gathered in the regions of maximum compression deformation). }
   \label{fig2}
   \end{center}
\end{figure}

As a working element in such radiometric sensors there is proposed a cubic monocrystal GaAs or CdTe with a surface (100) on which   the adsorbed atoms with an average concentration  ${{N}_{0d}}$ are located (figure~\ref{fig2}). The layer enriched with defects by adatoms is considered to be a ``film'' of $d$ thickness, $\rho$ density and Young's modulus $E$. This layer is rigidly connected with a substrate which is another  part of a monocrystal. The connecting plane of the film and the substrate is parallel to the surface (100). The \textit{z}-axis is directed towards the depth of the monocrystal and the axes \textit{x} and \textit{y}, along two orthogonal crystallographic directions of the type [100].

The surface acoustic wave propagating in the \textit{x}-axis direction with frequency $\omega (q,{{N}_{0d}})$ formes a dynamic deformation and interacts with the adsorbed atoms. The form of a deformed surface along the coordinate \textit{x} depending on time is described by the following function~\cite{Pel18}:

\begin{equation}
\label{eq1}
{{z}_{0}}(x,t)=\tilde{\varsigma }({{N}_{0d}})\cos (qx-\omega (q,{{N}_{0d}})t),
\end{equation}
where $\omega (q,{{N}_{0d}})={\omega }'(q,{{N}_{0d}})+\ri{\omega }''(q,{{N}_{0d}})$ is the  dispersion law of quasi-Rayleigh wave renormalized by adatoms; $q=2\piup /{{L}_{x}}$, ${{L}_{x}}$ is period of roughness (length of acoustic wave) along the \textit{x}-axis. It should be noted that the period of such roughness at a frequency of $\omega \sim $ 10 GHz is 540 nm.

The dispersion equation of a surface acoustic wave interacting  with the adsorbed atoms is as follows~\cite{Pel17}:

\begin{eqnarray}
 {(q^{2}+k_{t}^{2})^{2}-4q^{2}k_{t}k_{l}}&=&-\frac{2}{\beta}\frac{\omega ^{2}}{c_{l}^{2}}\frac{\theta _{d}N_{0d}}{{k}_{\text B}T\rho c_{l}^{2}}\cdot \frac{{D}_{d}{q}^{2}}{-\ri\omega +{{D}_{d}} \left( 1-\frac{2}{3}\frac{1-2\sigma }{K(1-\sigma )a}\frac{{{\theta }_{d}}^{2}}{{{k}_{\text B}}T}{{N}_{0d}} \right)q^{2}} \nonumber \\
  &\times& (1-l_{d}^{2}{{q}^{2}})\left[ {q}^{2} k_{t} \frac{\partial \tilde{F}}{\partial {{N}_{1d}}}+(q ^{2}+k_{t}^{2})\frac{{{\theta }_{d}}}{2a} \right],
  \label{eq2}
\end{eqnarray}
where $K$ is the modulus of elasticity; $l_d$ is the average of the square of a  characteristic length of interaction of adatoms with lattice atoms; $\sigma $ is the Poisson coefficient; $\theta_d$ is the surface deformation potential; ${{D}_{d}}$ is the diffusion coefficient of adatoms; $T$ is the temperature of a substrate; ${{k}_{\text B}}$ is the Boltzmann constant; ${{\xi }_{0}}$ is the solution of equation $f({{\xi }_{0}})=0$~\cite{Pel17}, which depends on the ratio between longitudinal ${{c}_{l}}$ and transversal ${{c}_{t}}$ speeds of the sound; $\beta =\frac{c_{t}^{2}}{c_{l}^{2}}$; ${k_{l,t}}^{2}=q^{2}-\frac{\omega ^{2}}{c_{l,t}^{2}}$;  $a$ is the lattice constant on the semiconductor surface; $\rho $ is the crystal density; ${{N}_{1d}}$ is the amplitude of periodic disturbance (${{{N}_{1d}}\ll{{N}_{0d}}}$)~\cite{Pel16}.

The real ${\omega }'(q,{{N}_{0d}})=\operatorname{Re}(\omega )$ and imaginary ${\omega }''(q,{{N}_{0d}})=\operatorname{Im}(\omega )$ parts of the expression $\omega (q,{{N}_{0d}})$ ultimately determine the correction to the dispersion law of the quasi-Rayleigh wave and its damping, and they are determined by relations (\ref{eq2}) and (\ref{eq3}). We take into account the nonlocal elastic interaction between the implanted impurity and the matrix atoms and the mirror image forces~\cite{Pel18}.
\begin{eqnarray}
 {\omega }'(q,{{N}_{0d}})={{c}_{t}}q{{\xi }_{0}}\Biggl\{  1-\frac{1}{{f}'({{\xi }_{0}})}\frac{2{{\xi }_{0}}{{\theta }_{d}}{{N}_{0d}}}{{{k}_{\text B}}T\rho c_{l}^{2}}\cdot \frac{{{D}_{d}}^{2}{{q}^{2}}[1-\frac{2}{3}\frac{1-2\sigma }{K(1-\sigma )a}\frac{{{\theta }_{d}}^{2}}{{{k}_{\text B}}T}{{N}_{0d}}]}{{{\left[ {{D}_{d}}q(1-\frac{2}{3}\frac{1-2\sigma }{K(1-\sigma )a}\frac{{{\theta }_{d}}^{2}}{{{k}_{\text B}}T}{{N}_{0d}}) \right]}^{2}}+c_{t}^{2}{{\xi }_{0}}^{2}} \nonumber  \\
  \text{ } \times (1-l_{d}^{2}{{q}^{2}})\left[ q\sqrt{1-{{\xi }_{0}}^{2}}\frac{\partial \tilde{F}}{\partial {{N}_{1d}}}+(2-{{\xi }_{0}}^{2})\frac{{{\theta }_{d}}}{2a} \right] \Biggr\} ,
\label{eq3}
\end{eqnarray}
\begin{eqnarray}
  {\omega }''(q,{{N}_{0d}})&=&\frac{1}{{f}'({{\xi }_{0}})}\frac{2\beta {{\xi }^{3}}{{\theta }_{d}}{{N}_{0d}}}{{{k}_{\text B}}T\rho }\cdot \frac{{{D}_{d}}{{q}^{2}}}{{{\left[ {{D}_{d}}q(1-\frac{2}{3}\frac{1-2\sigma }{K(1-\sigma )a}\frac{{{\theta }_{d}}^{2}}{{{k}_{\text B}}T}{{N}_{0d}}) \right]}^{2}}+c_{t}^{2}{{\xi }_{0}}^{2}}\nonumber \\
  &\times& (1-l_{d}^{2}{{q}^{2}})\left[ q\sqrt{1-{{\xi }_{0}}^{2}}\frac{\partial \tilde{F}}{\partial {{N}_{1d}}}+(2-{{\xi }_{0}}^{2})\frac{{{\theta }_{d}}}{2a} \right].
\label{eq4}
\end{eqnarray}

\section{The influence of the concentration of adsorbed atoms on the temperature coefficient of the resonant frequency}

SAW-sensors measuring the temperature on the surface of a monocrystal without adsorbed atoms are based on the change of the frequency of the quasi-Rayleigh wave depending on the temperature.
This effect is described by the temperature coefficient of the resonance frequency (TCF), which is defined as the relative change in the resonance frequency with temperature~\cite{Liu16}:

\begin{equation}
\label{eq5}
\alpha _{TCF}^{0}=\frac{1}{{{f}_{0}}}\frac{\text{d}{{f}_{0}}}{\operatorname{d}T}=\frac{1}{v}\frac{\text{d}v}{\text{d}T}-\frac{1}{\lambda }\frac{\text{d}\lambda }{\text{d}T},
\end{equation}
where ${{f}_{0}}$ is the resonance SAW frequency; $v$ is the velocity propagation of SAW; $\lambda$ is the SAW length; $T$ is the temperature of SAW-device.

The interaction between adatoms and a surface acoustic wave affects  the value of its resonant frequency, which is expressed by the temperature coefficient of frequency $\alpha _{TCF}$ . However, in expression~(\ref{eq5}) presented in~\cite{Liu16}, there is no component for the temperature coefficient of frequency that describes the influence of the adsorbed atoms concentration on the value of this coefficient.

In the case where a monocrystal surface has got adsorbed atoms, the change of the resonant frequency occurs not only due to the changes of temperature but also due to the presence of adatoms:

\begin{equation}
\label{eq6}
f({N}_{0d},T)=f_{0}(T)+\Delta f({N}_{0d},T).
\end{equation}

Let us perform logarithmic operations on two parts of the expression:

\begin{equation}
\label{eq7}
\ln f({N}_{0d},T)=\ln f_{0}(T)+\ln \left[ 1+\frac{\Delta f({N}_{0d},T)}{f_{0}(T)}\right].
\end{equation}

Taking into account  that ${\left[ \frac{\Delta f({N}_{0d},T)}{f_{0}(T) }\right] }^{2}\ll1$, we  receive:

\begin{equation}
\label{eq8}
\ln f({N}_{0d},T)=\ln f_{0}(T)+\frac{\Delta f({N}_{0d},T)}{f_{0}(T)}.
\end{equation}

We take the derivative of two parts of expression(\ref{eq8}). Taking into account the ratio  $f=\frac{v}{\lambda}$, we obtain:

\begin{equation}
\label{eq9}
\frac{1}{f}\frac{\text{d}f({N}_{0d},T)}{\text{d}T}=\frac{1}{v}\frac{\text{d}v }{\text{d}T}-\frac{1}{\lambda}\frac{\text{d}\lambda }{\text{d}T}+\left( \frac{1}{v}\frac{\text{d}\lambda }{\text{d}T}-\frac{\lambda}{v^2}\frac{\text{d}v }{\text{d}T}\right) \Delta f({N}_{0d},T)+\frac{\lambda}{v}\frac{\text{d}\Delta f({N}_{0d},T)}{\text{d}T}.
\end{equation}

Taking into account the concentration component, the dependence of the coefficient $\alpha _{TCF}$ on temperature and on the concentration of the adsorbed atoms is described by the following functional expression:

\begin{equation}
\label{eq10}
\alpha _{TCF}({N}_{0d},T)=\alpha _{TCF}^{0}(T)+\Delta \alpha _{TCF}({N}_{0d},T);
\end{equation}

\begin{equation}
\label{eq11}
\Delta \alpha _{\text{TCF}}({{N}_{0d}})=\frac{1}{v}\frac{\text{d}\lambda }{\text{d}T}\Delta f({{N}_{0d}})-\frac{\lambda }{{{v}^{2}}}\frac{\text{d}v}{\text{d}T}\Delta f({{N}_{0d}})+\frac{\lambda }{v}\frac{\text{d}\Delta f({{N}_{0d}})}{\text{d}T}.
\end{equation}

Taking into account the ratio ${\omega }'(q,{{N}_{0d}})-{\omega }'(q,0)=\Delta {\omega }'(q,{{N}_{0d}})=2\piup \Delta f(q,{{N}_{0d}})$ we get:

\begin{eqnarray}
\Delta f(q,{N}_{0d})&=&\frac{1}{{f}'({{\xi }_{0}})}\frac{{{\xi }_{0}}{{\theta }_{d}}{{N}_{0d}}}{\piup {{k}_{\text B}}T\rho c_{l}^{2}}\cdot \frac{{{D}_{d}}^{2}{{q}^{2}}(1-\frac{2}{3}\frac{1-2\sigma }{K(1-\sigma )a}\frac{{{\theta }_{d}}^{2}}{{{k}_{\text B}}T}{{N}_{0d}})(1-l_{d}^{2}{{q}^{2}})}{{{\left[ {{D}_{d}}q(1-\frac{2}{3}\frac{1-2\sigma }{K(1-\sigma )a}\frac{{{\theta }_{d}}^{2}}{{{k}_{\text B}}T}{{N}_{0d}}) \right]}^{2}}+c_{t}^{2}{{\xi }_{0}}^{2}}\nonumber  \\
  &\times& \left[ q\sqrt{1-{{\xi }_{0}}^{2}}\frac{\partial \tilde{F}}{\partial {{N}_{1d}}}+(2-{{\xi }_{0}}^{2})\frac{{{\theta }_{d}}}{2a} \right]. 
\label{eq12}
\end{eqnarray}

The function that determines the frequency of an acoustic oscillation vibration without the adsorbed atoms in the long-wave approximation $qa\ll 1$ has the following form~\cite{Dav76}:
\begin{equation}
\label{eq13}
\tilde{\omega }(q)=\frac{qa}{2}\sqrt{\frac{4{{E}_\text{ad}}Na}{{{m}_{1}}+{{m}_{2}}}},
\end{equation}
where $N$ is the number of cells; $m_1$, $m_2$ are the masses of atoms {Ga} and {As} respectively.

In the adiabatic approximation, Young's modulus ${{E}_\text{ad}}$ and the Poisson coefficient ${{\sigma }_\text{ad}}$ are described by the following relations~\cite{Lan70}:

\begin{equation}
\label{eq14}
{{E}_\text{ad}}=\frac{E}{1-ET{{\alpha }^{2}}/9{{C}_{p}}};  \quad     {{\sigma }_\text{ad}}=\frac{\sigma +ET{{\alpha }^{2}}/9{{C}_{p}}}{1-ET{{\alpha }^{2}}/9{{C}_{p}}}\,,
\end{equation}	
where $\alpha$ is the temperature coefficient of the bulk expansion;
${{C}_{p}}\sim \frac{12{{\piup }^{4}}}{5}{{N}_\text{V}}{{k}_{\text B}}{{\left( \frac{T}{{{\theta }_\text{D}}} \right)}^{3}}$ is the heat capacity at a constant pressure per unit volume; ${{\theta }_\text{D}}$ is the Debye temperature; ${{N}_\text{V}}$ is the number of atoms per unit volume.
The length of the surface acoustic wave is defined as $\lambda =\frac{2\piup v}{\tilde{\omega }(q)}$.  The velocity of SAW is described by the following relation~\cite{Lan70}:

\begin{equation}
\label{eq15}
v\simeq {{c}_{l}}=\sqrt{\frac{{{E}_\text{ad}}}{2\rho (1+{{\sigma }_\text{ad}})}\cdot \frac{2(1-{{\sigma }_\text{ad}})}{1-2{{\sigma }_\text{ad}}}}.
\end{equation}

The numerical calculation of the dependence of the dispersion law ${\omega }'\text{(}q\text{)}=\operatorname{Re}\omega \text{(}q\text{)}$ of a quasi-Rayleigh wave, the width ${\omega }''\text{(}q\text{)}=\operatorname{Im}\omega \text{(}q\text{)}$ of its phonon mode and the temperature coefficient of the resonant frequency depending on the concentration of the adsorbed atoms was performed for GaAs (100) semiconductor with the following parameter values~\cite{Pel18}:

$l_d=29$~\AA; $a=5.65$~\AA; $q=0.01$ \AA$^{-1}$; ${{c}_{l}}=4400$~m/s; ${{c}_{t}}=2475$~m/s; $\rho=5318$~kg/m$^{3}$; ${{\theta }_{d}}=$10~eV; $D_d=5\cdot {{10}^{16}}$ \AA$^{2}$/s; $\frac{\partial F}{\partial {{N}_{d}}}=0.1$~eV; $T=300$~K; $\sigma = 0.31$; $K = 0.47$~eV/\AA$^{3}$; $E = 0.54$~eV/\AA$^{3}$; $\alpha = 17.11\cdot{{10}^{-6}}$~K$^{-1}$; ${{\theta }_\text{D}}=360$~K; ${{N}_\text{V}}=4.42\cdot {{10}^{-2}}$~\AA$^{-3}$;  ${m}_{1}=69.72$~g/mol; ${m}_{2}=74.92$~g/mol.

\begin{figure}[!t]
   \begin{center}
   \includegraphics[width=95mm]{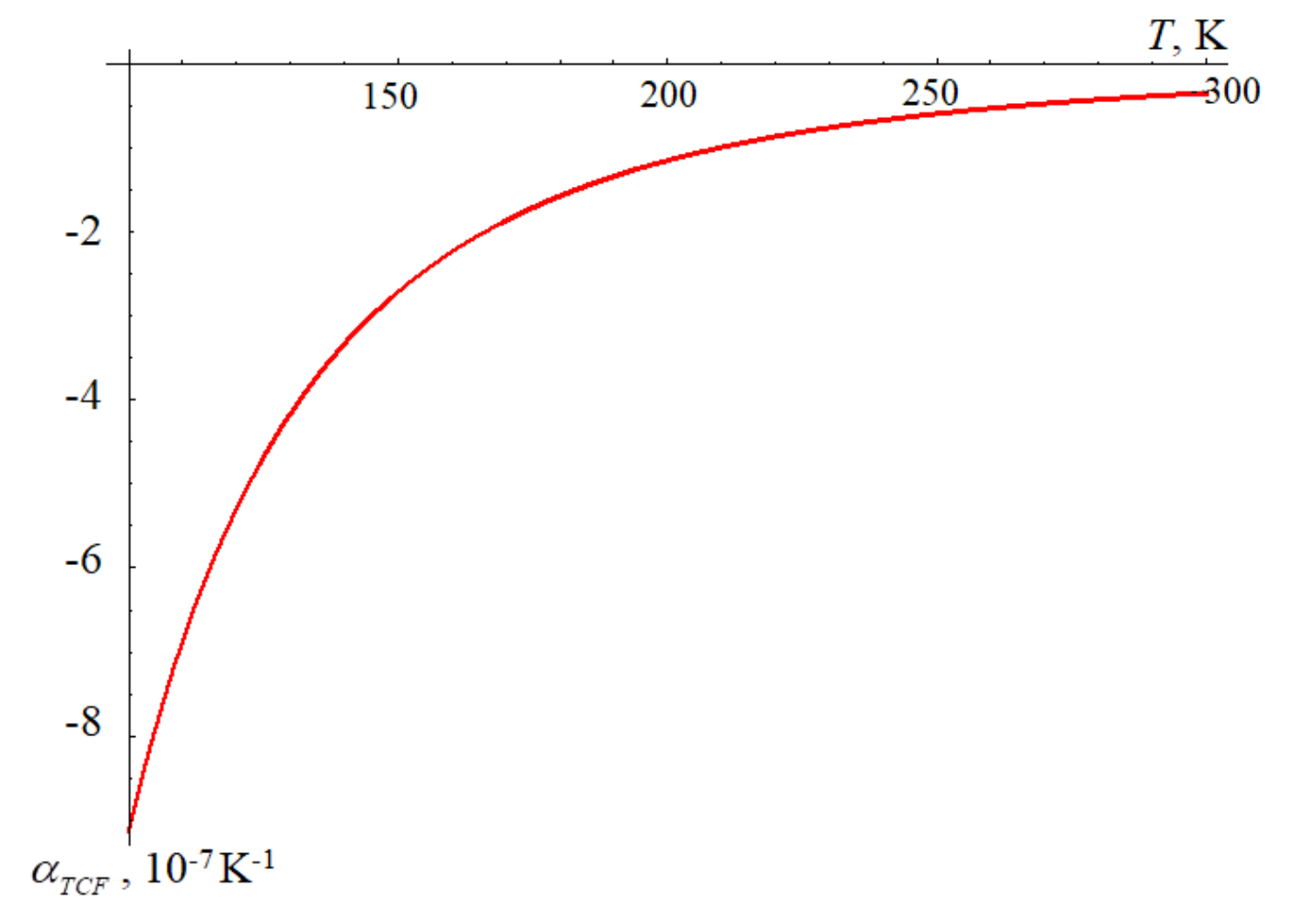}
   \caption{(Colour online) The dependence of the temperature coefficient of the resonant frequency $\alpha _{TCF}$ on the substrate temperature at the concentration of adatoms ${{N}_{0d}}=2\cdot {{10}^{13}}$ cm$^{-2}$}
   \label{fig3}
   \end{center}
\end{figure}

Figure~\ref{fig3} shows the graphical dependence of the temperature coefficient of the resonant frequency $\alpha _{TCF}$  of the surface acoustic wave on the substrate temperature at the concentration of adsorbed atoms $N_{0d}=2\cdot {{10}^{13}}$ cm$^{-2}$. We see that the absolute value of the temperature coefficient decreases with increasing temperature.

\begin{figure}[!t]
   \begin{center}
   \includegraphics[width=100mm]{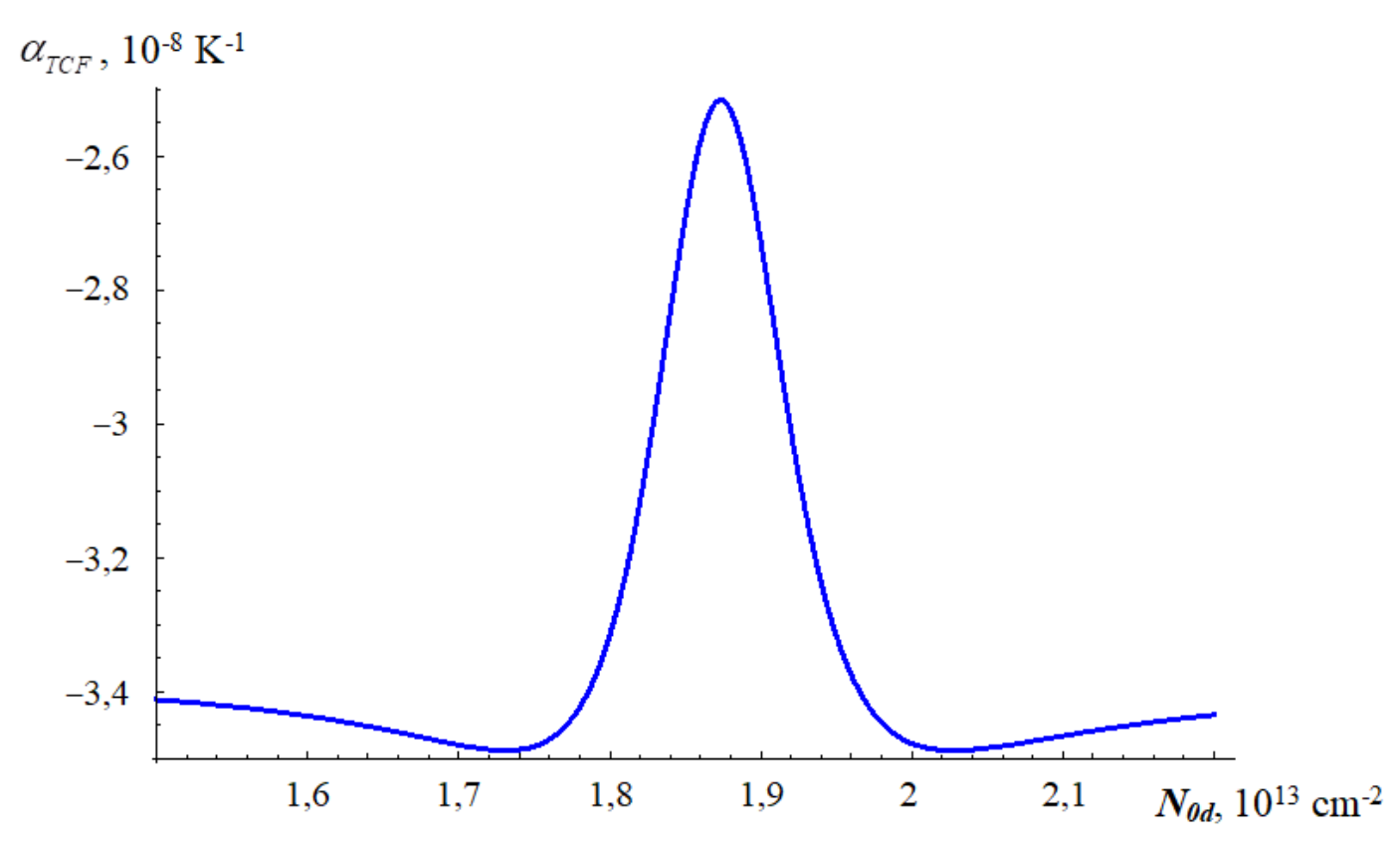}
   \caption{(Colour online) The dependence of the temperature coefficient of the resonant frequency $\alpha _{TCF}$ on the concentration of adsorbed atoms at a temperature 300 K. }
   \label{fig4}
   \end{center}
\end{figure}

The effect of the concentration of adsorbed atoms  on the temperature coefficient of the resonant frequency $\alpha _{TCF}$ at the temperature of 300 K is shown in figure~\ref{fig4}. We see that the functional dependence $\alpha _{TCF}=f({N}_{0d})$  has a nonmonotonous character with a maximum which corresponds to the adatoms concentration $N_{0d}=1.87\cdot {{10}^{13}}$ cm$^{-2}$.

The minima are at concentrations of  ${{N}_{0d}}=1.72\cdot {{10}^{13}}$ cm$^{-2}$ and  ${{N}_{0d}}=2.02\cdot {{10}^{13}}$ cm$^{-2}$. In the intervals of concentration ${{N}_{0d}}\leqslant 1.72\cdot {{10}^{13}}$ cm$^{-2}$ and $1.87\cdot {{10}^{13}}<{{N}_{0d}}\leqslant 2.02\cdot {{10}^{13}}$ cm$^{-2}$, there is a decrease of the graphical dependence $\alpha _{TCF}=f({N}_{0d})$.  In the intervals of concentration $1.72\cdot {{10}^{13}}<{{N}_{0d}}\leqslant 1.87\cdot {{10}^{13}}$ and ${{N}_{0d}}>2.02\cdot {{10}^{13}}$, the graphical dependence has an increasing character. Outside the studied range of concentrations, the graphical dependence is smoother.

This nonmonotonous character of the functional dependence $\alpha _{TCF}=f({N}_{0d})$ is due to two competing factors, i.e., the mechanical deformation of the surface crystal lattice created by the mechanical action of the adsorbed atoms (geometric scale factor), and the interaction of the adsorbed atoms through the deformation field created both by elastic acoustic quasi-Rayleigh wave and adsorbed atoms.

The results of calculations can be used to determine the concentration of adsorbed atoms at a known value of the temperature coefficient of the resonant frequency $\alpha _{TCF}$ of the surface acoustic wave using the proposed new class of radiometric SAW-sensors of temperature and the concentration of adsorbed atoms. A further research in this direction will make it possible to build the architecture of these radiometric sensors.

\section{Conclusion}

\begin{enumerate}
\item There was proposed a method for determining the concentration of adsorbed atoms using the radiometric sensors on surface acoustic waves on the basis of a model of a self-consistent connection of a quasi-Rayleigh wave with adsorbed atoms.
\item There was calculated the temperature-concentration  coefficient of the resonant frequency of the quasi-Rayleigh wave depending on the temperature and concentration of the adsorbed atoms, taking into account the theory of dispersion of the surface acoustic wave on the adsorbed surface of monocrystals having a Zinc blende structure.
\item It was shown that the functional dependence $\alpha _{TCF}=f(N{{}_{0d}})$ is of a non-monotonous character with a maximum corresponding to the adatoms concentration of ${{N}_{0d}}=1.87\cdot{{10}^{13}}$~cm$^{-2}$, which corresponds to the highest sensitivity of the radiometric sensor.
\end{enumerate}

\ukrainianpart

\title{Вплив концентрації адсорбованих атомів на температурний коефіцієнт резонансної частоти квазірелеївської хвилі}
\author{М.Я.~Сенета\refaddr{1,2},
        Р.М.~Пелещак\refaddr{1,2}, А.І.~Нестерівський\refaddr{1}, Н.І.~Лазурчак\refaddr{1},С.К.~Губа\refaddr{2}}
\addresses{
\addr{1} Дрогобицький державний педагогічний університет імені Івана Франка, вул. Івана Франка, 24, Дрогобич 82100, Україна
\addr{2} Національний університет ``Львівська політехніка'', вул. C.~Бандери, 12, Львів 79013, Україна
}
\makeukrtitle

\begin{abstract}
\tolerance=3000%
У роботі в межах моделі самоузгодженого зв’язку квазірелеївської хвилі з адсорбованими атомами запропоновано спосіб побудови нового класy радіометричних сенсорів температури та концентрації адсорбованих атомів на поверхневих акустичних хвилях. На основі розвинутої теорії дисперсії і ширини акустичної моди квазірелеївської хвилі на адсорбованій поверхні монокристалів із структурою цинкової обманки розраховано температурний коефіцієнт резонансної частоти поверхневої акустичної хвилі в залежності від температури і концентрації адсорбованих атомів.
\keywords поверхнева акустична хвиля, адсорбовані атоми, температурний коефіцієнт резонансної частоти

\end{abstract}

\end{document}